# Spatial Cluster-based Copula Model to Interpolate Skewed Conditional Spatial Random Field

Debjoy Thakur[1], Ishapathik Das[2], Shubhashree Chakravarty [3]

[1,2]Department of Mathematics and Statistics, Indian Institute Of Technology, Tirupati.
[3]Machine Intelligence Unit, Indian Statistical Institute, Kolkata.

**Abstract**

Interpolating a skewed conditional spatial random field with missing data is cumbersome in the absence of Gaussianity assumptions. Maintaining spatial homogeneity and continuity around the observed random spatial point is also challenging, especially when interpolating along a spatial surface, focusing on the boundary points as a neighborhood. Otherwise, the point far away from one may appear the closest to another. As a result, importing the hierarchical clustering concept on the spatial random field is as convenient as developing the copula with the interface of the Expectation-Maximization algorithm and concurrently utilizing the idea of the Bayesian framework. This paper introduces a spatial cluster-based C-vine copula and a modified Gaussian kernel to derive a novel spatial probability distribution. Another investigation in this paper uses an algorithm in conjunction with a different parameter estimation technique to make spatial-based copula interpolation more compatible and efficient. We apply the proposed spatial interpolation approach to the air pollution of Delhi as a crucial circumstantial study to demonstrate this newly developed novel spatial estimation technique.

**Keywords—** Von-Mises distribution, Expectation-Maximization algorithm, Hierarchical Spatial Clustering, Spatial Copula Interpolation, Bayesian Spatial Copula Interpolation

## 1 Introduction

The upward trend of Particulate Matter ($PM$) concentrations in the atmosphere and air pollution has become the greatest threat to human civilization daily. Every year, nearly 0.8 million people die due to the direct and indirect effects of air pollution, and approximately 4.6 million people suffer from serious diseases such as chronic obstructive pulmonary disease (COPD), respiratory hazards, premature deaths, and so on [ (Auerbach, 2012), (Lim, 2012), (WHO, 2016)]. It is unavoidable that the air-pollutant concentration is estimated with greater accuracy to control air pollution. The spatial and spatio-temporal application of geostatistics is crucial during prediction.

A very interesting task in geostatistics is, to interpolate a value of the target variable at a specific time stand, in an unobserved location, taking its surroundings into account. In this scenario, the researchers prefer to use Inverse Distance Weight (IDW), Ordinary Kriging (OK), Universal Kriging (UK), Disjunctive Kriging (DK), etc [ (Cressie, 2015), (Isaaks, 1989)]. Because of significant advances in data science, many scientists prefer neural networking-based spatial and spatio-temporal interpolation techniques like Geo-Long Short Time Memory (Geo-LSTM), Random Forest Regression Kriging (RFK), and others [ (Ma, 2019), (Shao, 2020)]. The previously mentioned algorithms use the variance-covariance function as a measure of dependence. The main drawback of this traditional spatial interpolation algorithm is the gaussianity assumption, rarely met. The neural networking-based algorithms outperform, but the mathematical justification is difficult; as a result, applying this model in other cases can be challenging. These significant limitations promote the use of the copula-based spatial and spatio-temporal interpolation approach. This copula-based spatial interpolation technique is both theoretically and practically flexible. Spatial variability is easily captured using copula, and then spatial lag-based Gaussian and non-Gaussian bi-variate copulas are created to interpolate four different groundwater quality parameters in Baden-Wurttemberg [ (Bardossy A. , 2006)]. Following that, many advances in spatial copula are established, for example, the utilization of asymmetric copulas to measure spatial independence [ (Bardossy A. a., 2009)], the use of Gaussian and non-Gaussian vine copula to derive the conditional distribution in the unobserved location [ (Bardossy A. , 2011)], employing the convex combination of archimedean copulas to kriging [ (Sohrabian, 2021)]. Besides these, application of the Gaussian Copula (GC) via bayesian framework to predict the maximum temperatures in the Extremadura Region in southwestern Spain, for the period 1980–2015 [ (Garcia, 2021)], and making use of a bi-variate copula [ (Masseran, 2021)]measures the dependence between air pollution severity and duration. Employing the copula-based bias-correction method [ (Alidoost, Stein, Su, & Sharifi, 2021)] develops three multivariate copula based quantile regression to map daily air temperature data, modeling the spatial pattern and spatial dependence structure of different climate variables, for instance, precipitation, air temperature by using regular (C- and D-) Vine copula and the student t-copula [ (Khan, Spöck, & Pilz, 2020)]. Utilizing extreme value models like, the generalized Pareto distribution (GPD), [ (Masseran & Hussain, 2020)] proposes a combination of a GPD model with copulas to establish a dependence model between $PM_{10}$ and a set of four major pollutant variables, namely, CO, $NO_2$, $SO_2$, and $O_3$, employing the concept of an extra-parameterized multivariate extreme valued copula [ (Carreau & Toulemonde, 2020)] introduces a spatial copula model, application of D-vine copula based quantile regression to model the spatial and Spatio-temporal data of COVID-19 in Italy [ (D'Urso, De Giovanni, & Vitale, 2022)]. After capturing the seasonality and temporal dependency of the daily mean temperature data set a new spatial distance based R-vine copula is introduced [ (Erhardt, Czado, & Schepsmeier, 2015)], and the usage of spatial vine copula based upon the spatial lag [ (Gräler, 2014), (Bostan, Stein, Alidoost, & Osei, 2021)]. With the help of the Metropolis-Hastings Algorithm (MHA), they have improved spatial copula in the Bayesian framework to approximate the posterior predictive density whereas, this method is limited to the Gaussian copula family [ (Kazianka & Pilz, 2011)]. Utilizing spatial vine copula and dimension reduction transformation [ (Musafer & Thompson, 2017)] creates a non-linear optimal multivariate spatial design to mitigate prediction uncertainty of more than one variable. Introducing the copula-based semi-parametric algorithm [ (Quessy, Rivest, & Toupin, 2015)] models the stationary and isotropic spatial random fields. Considering the concept of generalized method of moments [ (Bai, Kang, & Song, 2014)] proposes a pairwise composite likelihood with the help of pair copula. Discovering a spatial factor copula model [ (Krupskii, Huser, & Genton, 2018)] combines the flexibility of a copula, the accountability of factor models, and the tractability of GC in higher dimensions to fit spatial data at different temporal replicates. Extending the spatial GC interpolation method [ (Gnann, Allmendinger, Haslauer, & Bárdossy, 2018)] predict the primary variable, groundwater quality using the categorical information of the primary variable considered as a secondary variable. Introducing two spatial copula interpolation methods, accountable to justify the spatial dependency of an air temperature of a particular location on its geo-spatial neighborhood, and another mixture copula [ (Alidoost, Stein, & Su, 2018)] explain the relationship of a variable with other covariates. A translation process (TP) is discovered [ (Richardson, 2021)] for a non-Gaussian spatial copula interpolation process and too effective to model, where the link function is not well defined. [ (Wang, Wang, Deng, Zou, & Wang, 2021)] develops a spatio-temporal heterogeneous copula-based kriging (HSTCAK). The space-time dependency is measured by the copula function and the spatial variability with temporal similarity and fuzzy clustering employed for partition to mitigate the heterogeneity problem. Introducing crucial advancements in the spatial copula approach like tail dependency and asymmetric dependency and extension of the linear model of coregionalization specifically for modeling the multivariate spatial data [ (Krupskii & Genton, 2019)], and they use cross-covariance function as the measure of spatial dependence.

The research articles used copulas in the spatial interpolation very well in the literature, but there are some constraints that the previous authors have ignored. To estimate parameters, they use the Maximum Likelihood estimate, which does not provide a good estimate in presence of missing data. (ii) After creating spatial copula interpolation, they fix one point and calculate the probability distribution at different lags from that point. As a result, the copula is limited within the fixed reference frame, but the reference frame is random in reality. That is why, we consider the random points and form a cluster, based on relative distance. (iii) At the time of spatial clustering, they disregard the significance of creating the disjoint regions, thus the intersection part is the most affected area, where the different effects of different clusters become confused with each other (iv) They use conditional expectation for interpolation, but it is invalid for the extremely valued probability density function (PDF).

In this study we evolve a novel spatial cluster-based copula modeling in different frameworks. We devide the entire spatial domain into k spatial clusters to get m number of spatial regions i.e. $\mathcal{L}_i \subseteq \mathbf{R}^{2 \times 2}$ which is the class of all possible set of points in a spatial region. We create a conditional spatial random field $Y: \mathcal{L}_i \times \mathcal{F}_{\mu^*} \to \mathcal{M}$. Here, $\mathcal{F}_{\mu^*}$ is an induced probability space created using caratheodory's extension theorem and $\mathcal{F}_{\mu^*} = \{A \in \mathcal{L}_i \mid A \text{ is } \mu^* - \text{measurable } i.e. \mu^*(A) \leq \delta\}$ and $Y$ is $< \mathcal{L}_i \times \mathcal{F}_{\mu^*}, \mathcal{M} >$ measurable random field and $\mathcal{M} \subseteq \mathbf{R}$. Our objective in this research is to predict $Y$ at an unobserved location on $s_0 \in \mathcal{L}_i$ based upon the n distinct observed location $s_1, s_2, \ldots, s_n \in \mathcal{L}_i$ using spatial copula interpolation algorithm in classical and bayesian framework.

The outline of this study is as follows: the details of the algorithm are introduced in Section (2), the study area and the behavior of data are described in Section (3), the results and discussion regarding the case study are summarized in Section (4) and the conclusion is made in Section (5).

## 2 Method

### 2.1 Fitting Marginal Distribution

In this particular section, we illustrate how to fit the ideal univariate parametric distribution on the empirical marginal likelihood distribution. We have divided the complete appropriate procedure into a few steps: (i) Choice of a family of distributions, (ii) Some sort of suitable marginal probability distribution among that family, and (iii) Typically the estimate of the parameter of these suited marginal probability distribution. For step (i) we use the Cullen and Frey graph of skewness-kurtosis plot, for step (ii) we utilize Kernel Density Estimation (KDE) centering on Akaike Information Criteria (AIC), and Bayesian Information Criteria (BIC), Log-likelihood value (LogLik), and Kolmogrov Smirnov (KS) values. Although we face a real challenge in coordination (iii) because there is missing data so the Maximum Likelihood Estimation (MLE) of the parameter is not recommendable. Therefore, for the Parametric Exponential Family distribution (PEF) we are making the use of Expectation-Maximization algorithm (EM) [ (McLachlan & Krishnan, 2007)] and for the circular probability distributions, we use Uniformly Minimum Variance Unbiased Estimator (UMVUE) technique. For PEF we consider the fitted distribution is Log-Normal (LN) probability distribution then, $logW \sim \mathcal{N}(\mu, \sigma^2)$. Let, $w_i; i = 1,2,3,\ldots,n_1$ are the observed data points and $w_i; i = n_1 + 1, n_1 + 2, n_1 + 3, \ldots, n_2$ are the un-observed data points. The likelihood function of $(\mu, \sigma)$ based upon the observed data:

$$logL_o(\mu, \sigma) = \frac{-1}{2\sigma^2} \cdot \sum_{i=1}^{n_1} (logw_i - \mu)^2 \\ -\sigma \cdot \sqrt{2\pi} \sum_{i=1}^{n_1} logw_i \tag{1}$$

The complete, observed and missing data vectors are respectively, $\mathbf{x} = (w_1, w_2, w_3, \ldots, w_{n_2})^T$ ; $\mathbf{y} = (w_1, w_2, w_3, \ldots, w_{n_1})^T$ and $\mathbf{z} = (w_{n_1+1}, w_{n_1+2}, w_{n_1+3}, \ldots, w_{n_2})^T$ reveals $\mathbf{x} = \mathbf{y} \cup \mathbf{z}$. The complete data log-likelihood function is:

$$logL_c(\mu, \sigma) = \frac{-1}{2\sigma^2} \cdot \sum_{i=1}^{n_2} (logw_i - \mu)^2 \\ -\sigma \cdot \sqrt{2\pi} \sum_{i=1}^{n_2} logw_i \tag{2}$$

Let's consider the E-step on the $(m+1)^{th}$ iteration of the EM algorithm where $(\mu^{(m)}, \sigma^{(m)})$ is the value after the $m^{th}$ iteration of EM. Using the Equation (2) we compute the conditional expectation of log-likelihood of the complete data (CElikC) based upon the updated value at the $m^{th}$ iteration, defined as $\mathcal{Q}\left((\mu, \sigma) | (\mu^{(m)}, \sigma^{(m)})\right)$ in the following:

$$\begin{aligned} &\mathcal{Q}\left((\mu, \sigma) | (\mu^{(m)}, \sigma^{(m)})\right) \\ &= E_{(\mu^{(m)}, \sigma^{(m)})}[logL_c(\mu, \sigma) | \mathbf{y}] \\ &= \int_{\mathbf{z}} (logL_c(\mu, \sigma) | \mathbf{y}) \cdot f(\mathbf{z}|\mathbf{y}, (\mu^{(m)}, \sigma^{(m)}) d\mathbf{z} \\ &= \int_{\mathbf{z}} \frac{(logL_c(\mu,\sigma)|\mathbf{y})}{f(\mathbf{y};(\mu^{(m)},\sigma^{(m)}))}) \cdot f(\mathbf{z}, \mathbf{y}; (\mu^{(m)}, \sigma^{(m)}) d\mathbf{z} \\ &\leq E_{(\mu^{(m)}, \sigma^{(m)})} \left[\frac{logL_c(\mu,\sigma)}{logL_o(\mu^{(m)},\sigma^{(m)})}\right] \end{aligned} \tag{3}$$

Using the Equation (1), (2) we will simplify the Equation (3) and then in M-step we maximize $\mathcal{Q}\left((\mu, \sigma) | (\mu^{(m)}, \sigma^{(m)})\right)$. Therefore the updated values are $(\mu^{(m+1)}, \sigma^{(m+1)})$ which is defined in the following:

$$(\mu^{(m+1)}, \sigma^{(m+1)}) = argmax_{(\mu,\sigma)} \mathcal{Q}\left((\mu, \sigma) | (\mu^{(m)}, \sigma^{(m)})\right) \tag{4}$$

Utilizing Equation (4), we get the estimate of the parameter. But to estimate the parameter of a circular probability distribution for example, Von-Mises (VM) distribution, avoiding the computational complexity of EM algorithm, absence of closed form, and due to the presence of Bessel Function ($I_n(k)$) we introduce a new theorem regarding the completeness and sufficiency of an estimator to deduce a UMVUE of the parameter of VM distribution in the following:

**Theorem 2.1** *If $X_i \sim^{iid}$ VM then $\frac{I_0(k) \cdot \cos(x_i)}{I_1(k)}$ and $\frac{I_0(k) \cdot \sin(x_i)}{I_1(k)}$ are the UMVUE of $\cos\mu$ and $\sin\mu$ respectively and their corresponding variances are*

$$\begin{aligned} var(\cos(x_i)) &= \frac{1}{2} + \frac{I_2(k) \cdot \cos(2\mu)}{2I_0(k)} - \left(\frac{I_1(k) \cdot \cos(\mu)}{I_0(k)}\right)^2 \\ var(\sin(x_i)) &= \frac{1}{2} - \frac{I_2(k) \cdot \sin(2\mu)}{2I_0(k)} - \left(\frac{I_1(k) \cdot \sin(\mu)}{I_0(k)}\right)^2 \end{aligned} \qquad (5)$$

*Proof.* See Appendix (6.1)

Using Theorem (2.1) and trigonometric inverse function we get the initial value and, update the parameter values of VM distribution like LN (see Appendix (6.2)).

### 2.2 Copula

A copula is used to model the dependence between two or more random variables, for formulating the joint multivariate distribution from the marginal cumulative distribution function (CDF). Let, $X_1, X_2, \ldots, X_n$ be $n$ random variables (RV) with corresponding marginal CDF s are respectively, $F_1(x_1), F_2(x_2), \ldots, F_n(x_n)$. The joint distribution function can be defined as, $F_{X_1,X_2,\ldots,X_n}(x_1, x_2, \ldots, x_n)$ which is the product of marginal and conditional distribution but, because of the complexity of this approach with the increasing number of random variables this approach is not applicable for the large number of random variables. Therefore, the copula function is defined to create a multivariate distribution from the $n$ marginal distribution [ (Nelsen, 2007), (Sklar, 1973)] to model the dependence between the multidimensional variables in the following way:

$$C: [0,1]^n \to [0,1]$$

$$F_{X_1,X_2,\ldots,X_n}(x_1, x_2, \ldots, x_n) = C(F_1(X_1), F_2(X_2), \ldots, F_n(X_n)). \qquad (6)$$

There are different types of copula families, for example, Gaussian, Archimedean, Product, etc and they behave in a different manner in the tail part of the distribution. Compared to the other traditional multivariate, elliptical, Archimedean copulas, and vine copulas (VC) are more flexible to capture the inherent dependency. Under some certain regularity conditions it's possible to express the $n-$dimensional multivariate copula mentioned in the Equation (6) as multiplication of pair-copulas [ (Aas, Czado, Frigessi, & Bakken, 2009)] in the following iterative approach. For $n = 2$ the bi-variate probability density function (BPDF) is:

$$f(x_1, x_2) = f_{12}(x_1|x_2) \cdot f_2(x_2) = c_{12}(F_1(x_1), F_2(x_2)) \cdot f_1(x_1) \cdot f_2(x_2). \qquad (7)$$

In the Equation (7) $c_{12}(.,.)$ is the applicable pair-copula density function (PCDF) for $F_1(x_1)$ and $F_2(x_2)$. For $n = 3$ the tri-variate probability density function (TPDF) is:

$$f(x_1, x_2, x_3) = f_{123}(x_1|x_2, x_3) \cdot f_{23}(x_2|x_3) \cdot f_3(x_3)$$
$$= c_{12}(F_1(x_1), F_2(x_2)) \cdot c_{23}(F_2(x_2), F_3(x_3)) \qquad (8)$$
$$\cdot c_{13|2}(F(x_1|x_2), F(x_3|x_2)) \cdot f_1(x_1) \cdot f_2(x_2) \cdot f_3(x_3).$$

In Equation (8) $c_{23}(.,.), c_{13|2}(.,..)$ are the applicable PCDF and conditional PCDF (CPCDF) respectively. Likewise, for $n = 4$ the four-variate probability density function (FPDF) is:

$$\begin{aligned}f(x_1, x_2, x_3, x_4) &= f_{4123}(x_4|x_1, x_2, x_3) \cdot f_{312}(x_3|x_1, x_2) \cdot \\ & f_{21}(x_2|x_1) \cdot f_1(x_1) = c_{12}(F_1(x_1), F_2(x_2)) \cdot \\ & c_{13}(F_1(x_1), F_3(x_3)) \cdot c_{14}(F_1(x_1), F_4(x_4)) \cdot \\ & c_{23|1}(F(x_2|x_1), F(x_3|x_1)) \cdot c_{24|1}(F(x_2|x_1), F(x_4|x_1)) \cdot \\ & c_{34|2}(F(x_3|x_2), F(x_4|x_2)) \cdot f_1(x_1) \cdot f_2(x_2) \cdot f_3(x_3) \cdot f_4(x_4)\end{aligned} \qquad (9)$$

VC is a graphical approach representing an $n$-dimensional multivariate PDF (MPDF) using $\frac{n.(n-1)}{2}$ suitable PCDF in a hierarchical manner where the dependence structures of $(n-1)$ have unconditional PCDF and that of the remaining has conditional PCPDF (CPCDF). In this paper, we focus on C-Vine Copula (C-VC), because of its better flexibility.

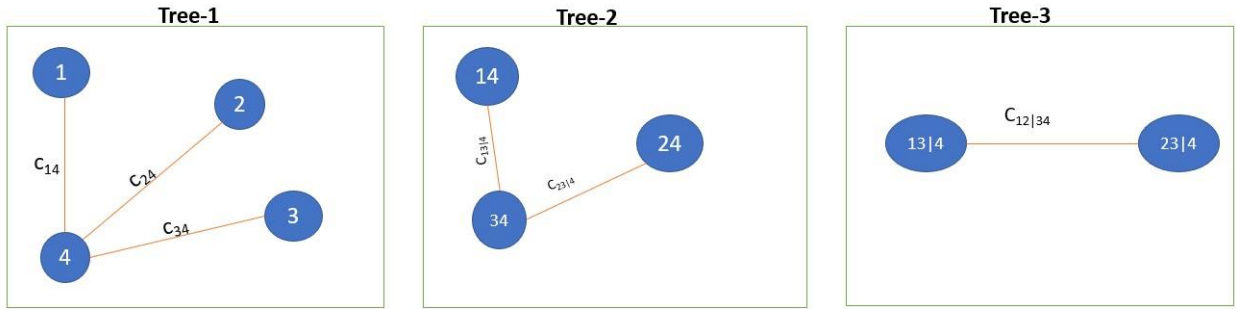

**Figure 1:** Detail of the tree of the Vine Copula.

A C-VC with 4-variables has 3 trees, $T_j$ and each tree, $T_j$ has $4 - j + 1$ nodes and $4 - j$ edges where $j = 1,2,3$ like Figure (1). In tree $T_1$ each edge between two nodes represent the PCDF. From Figure (1) in tree $T_2$ the edges between each node is CPCDF where $c_{13|4}$ denotes the CPCDF of the first and third variable given the fourth variable and $c_{23|4}$ represents the CPCDF of the second and third variable given the fourth variable. In the tree $T_3$ each node is connected by an edge representing CPCDF (Figure (1) ) of the first and second variable given third and fourth variable where $C_{12|34} = F(x_1, x_2|x_3, x_4)$.

### 2.3 Spatial Interpolation

In this section we propose two novel spatial interpolation approaches combining spatial clustering, knowledge of copula and C-VC assuming the directional stationarity of data is defined in the following:

#### 2.3.1 Spatial Copula Estimation

Let $S$ be the spatial domain of interest for the spatial interpolation purpose. This spatial clustering technique is established upon the distance and degree of similarity between two spatial points. Hierarchical spatial clustering (HSC) is performed using the complete linkage method [ (Hubert, 1974)], cutoff values of Haversine distance (HD) [ (Gade, 2010)] and auto-correlation between each pair of points. Here, the number of HSC and HSC's radius are the two most important parameters. According to the principle of HSC, it is obvious that the sum of squares within a cluster (SSW) should be lesser than the sum of squares between the clusters (SSB). Making use of the Elbow method while the SSW reaches a plateau, we will consider that is the optimal number of HSC and to determine HSC's radius arranging HSC's height in ascending order, that considers a significant height as a HSC's radius. Therefore, in this context, we can think of the HSC as a spatial field defined in the following Equation (10)

$$N_i = \{(ob_{i_1}, ob_{i_2}): HD(ob_{i_1}, ob_{i_2}) \leq HD_{cut} \ \& \ \rho(||ob_{i_1} - ob_{i_2}||) \geq r_{cut} \ \& \ i_1 \neq i_2\}$$
$$\cup \ \{(ob_{i_j}, y_{ij}): HD(ob_{i_j}, y_{ij}) \leq HD_{cut} \ \& \ \rho(||ob_{i_j} - y_{ij}||) \geq r_{cut}\} \quad (10)$$

Let $N_1, N_2, \ldots, N_k$ be the $k$ clusters, $y_{ij}$ be the $j^{th}$ unobserved point, and $ob_{i_j}$ be the $j^{th}$ observed point of the $i^{th}$ HSC where, $j = 1,2, \ldots, n_i$; $i = 1,2,3, \ldots, k$ where $n_i$ be the number of the unobserved points in $i^{th}$ HSC and $c_i$, the centre of $i^{th}$ cluster. A threeshold HD ($HD_{cut}$) is chosen $\ni dist(c_i, y_{ij}) \leq HD_{cut}$ that is surrounding the centre of each HSC, a circle is constructed with radius of $HD_{cut}$ units and $r_{cut}$ refers the spatial auto correlation cutoff to maintain the spatial continuity of HSC. For $k$ clusters maximum number of distinct spatial regions (SR) is $2^k - 1$. Let $\vec{v}$ be the presence vector of all unobserved points $y_{ij} = (lon, lat)$, where $lon$ and $lat$ stand for longitude and latitude of an unobserved point respectively. Now, we create a linear map in the following:

$$f: \mathcal{R}^{2\times 1} \to \mathcal{R}^{k\times 1} \Rightarrow f(y_{ij}) = \vec{v} \quad (11)$$

In the Equation (11) $\vec{v}$ is a binary vector of length $k$, where $\vec{v} = [v_1, v_2, \ldots, v_k]$

$$v_l = \begin{cases} 1 & if \ (lon, lat) \in C_l \\ 0 & otherwise \end{cases}$$

where $l = 1,2,3, \ldots, k$. The maximum number of presence vectors for $k$ clusters is $2^k - 1$. As a result, we obtain at most $2^k - 1$ distinct SR in our entire study area. Let $R_1, R_2, \ldots, R_m$ be the $m$ SR where $m \leq 2^k - 1$. Let $v_i = 0 \ \forall \ i = 1,2, \ldots, (i-1), (i+1), \ldots, k$ and $v_i = 1$ that denotes $y_{ij}$ is inside $N_i$ only.

**(a)**

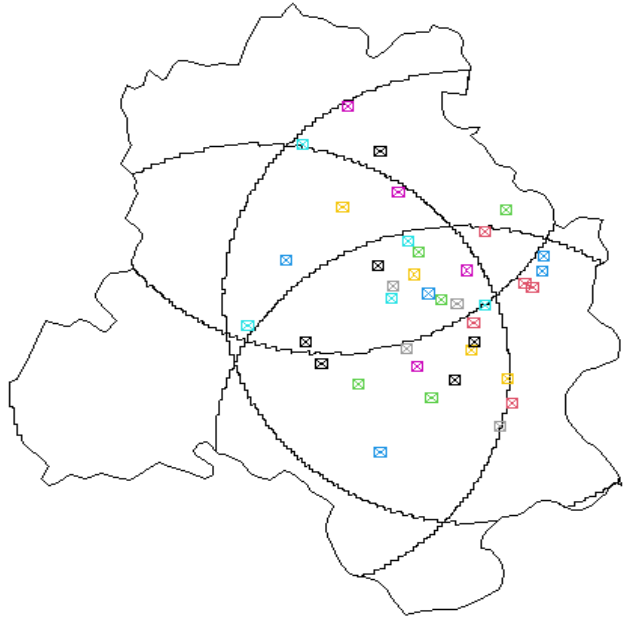

**(b)** **Clustering of the Spatial Field**

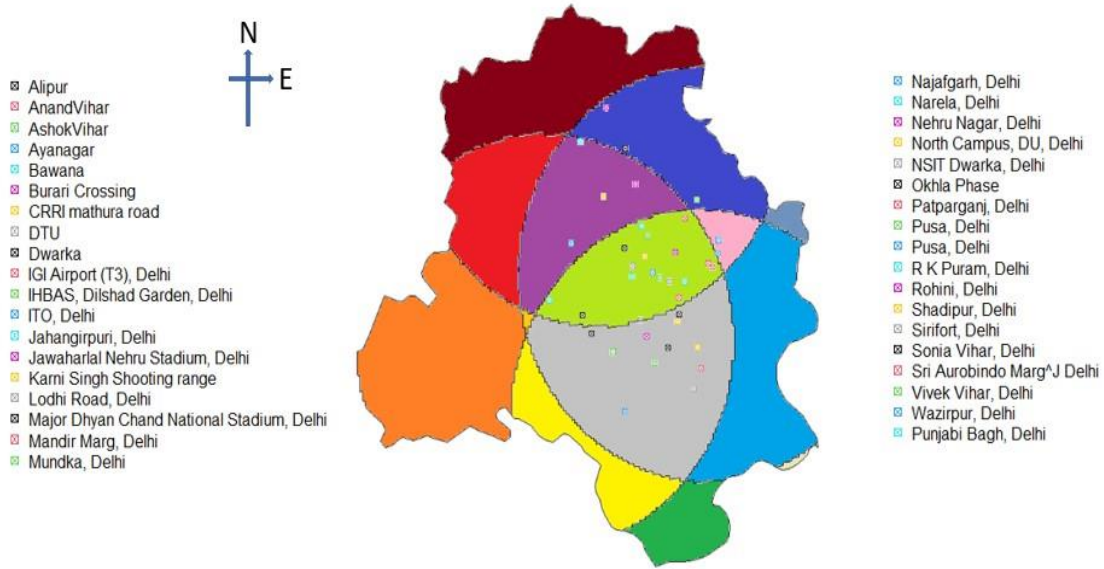

**Figure 2: (a)** The entire spatial domain, Delhi is divided into four clusters and the corresponding monitoring stations.

**(b)** The entire geographical area is split up into some disjoint regions which are shaded by different colour and the corresponding observed points belonging into each region.

In Figure (2 a) we divide the entire spatial domain into some HSC grounded on the HD of monitoring stations and degree of homogeneity where each circle denotes a HSC, and the dotted points represent the observed monitoring stations contained in that HSC. As a result, the whole area is split into some disjoint SR and the dotted points represent the monitoring stations, included in that SR (Figure (2 b)) are salient points to interpolate along the surface of each SR.

Utilizing the concept of copula, we transform a HSC to a spatial random field (SRF) to predict the values on the unobserved location. Therefore, we first concentrate on the univariate marginal probability distribution (MDF) of inclusion of the latitude ($\omega_i$) and longitude ($\beta_i$) of an observed location being included in $R_i$ separately. The corresponding MDF are respectively $P(\{\omega: \omega \in R_i\})$ and $P(\{\beta: \beta \in R_i\})$. Using the concept of copula, we have created the Bi-variate probability distribution function (BDF) of inclusion of latitude and longitude in the $i^{th}$ SR, and Kendall's $\tau$ is considered as the measure of association between two RVs. So, the joint BDF is as follows:

$H(x_1, x_2) = C(F_\omega(X_{1\omega}), F_\beta(X_{2\beta}))$ where $C: [0,1]^2 \rightarrow [0,1]$.

To evaluate the CDF of that spatial random process (SRP), a composite function of two RVs i.e.,

$Y(.,.) \equiv \{Y(\omega, \beta): \omega \in R_i, \beta \in R_i\}$ we implement the KDE to get the MDF of $Y$ i.e., $F(y)$ and making use of copula we deduce the joint Tri-variate probability distribution (TDF) as follows:

$H(x_1, x_2, y) = C(F_\omega(X_{1\omega}), F_\beta(X_{2\beta}), F(Y))$ where $C: [0,1]^3 \rightarrow [0,1]$.

After getting, the BDF ($H(x_1, x_2)$) and TDF ($H(x_1, x_2, y)$) using copula we find out the conditional PDF (CPDF) defined in the following two equations:

$$f(y_1 | x_1, x_2) = \frac{\frac{\partial^3 H(x_1, x_2, y)}{\partial x_1 \partial x_2 \partial y}}{\frac{\partial^2 H(x_1, x_2)}{\partial x_1 \partial x_2}} \quad (12)$$

Here, we assume two RVs, latitude ($X_1$) and longitude ($X_2$) follow Uniform[$a_1, b_1$] and Uniform[$a_2, b_2$] respectively. Now we generate random points ($x_1, x_2$) along, $S$ to measure the CDF of SRP i.e., $Y(x_1, x_2)$ having different CDF for each geographical position. Since, there are some missing data, we use EM algorithm to estimate the parameter at the time of fitting MDF. Applying copula we get the joint TDF of $Y(x_1, x_2), X_1, X_2$ is defined as $F(y, x_1, x_2)$. Next, we will introduce the HSC-based SI that Spatial Copula interpolation (SC). Therefore, we split $S$ in $m$ number of regions making use of the HSC algorithm, discussed earlier. Let's consider, $R_1$ has three observed locations $ob_1, ob_2, ob_3$. In that region we can generate a number of gridded points not necessarily of uniform size, out of those unobserved points we consider one unobserved point, defined as $un_{R_1}^j$ which is the $j^{th}$ point in $R_1$. Applying the conditional copula (from Equation (12)) we establish the conditional copula based probability distribution function (CCDF) for each un-observed point in a SR i.e. $F_{un_{R_1}^j}(y)$.

$$F_{un_{R_1}^j}(y) = P[Y(x_{1, un_{R_1}^j}, x_{2, un_{R_1}^j}) \leq y | X_1 = x_{1, un_{R_1}^j}, X_2 = x_{2, un_{R_1}^j}] \quad (13)$$

From the Equation (13) we get the CCDF of $j^{th}$ un-observed point included in the first SR. That lets us calculate the CCDF of SRP, $Y$ at the unobserved centroid of a cluster and making the use of CCDF we can calculate conditional copula based probability density function (CCPDF). The mathematical formulation is described in the following:

$F_{un_{R_1}^j}(y) \quad = \sum_{i \in R_1} \alpha_{ij} \cdot P[Y(x_{1, ob_i}, x_{2, ob_i}) \leq y | X_1 = x_{1, ob_i}, X_2 = x_{2, ob_i}]$

$\Rightarrow f_{un_{R_1}^j}(y) \quad = \sum_{i \in R_1} \alpha_{ij} \cdot f_{ob_i}(y | x_1, x_2)$

$\Rightarrow argmax_y f_{un_{R_1}^j}(y) \quad = \sum_{i \in R_1} \alpha_{ij} \cdot argmax_y f_{ob_i}(y | x_1, x_2)$

(14)

In the Equation (14) the weights are defined as $\alpha_{ij}$. This weights are the proportional to the spatial auto correlation function (ACF) but inversely proportional to the degree of separation. So the required $\alpha_{ij}$ is defined in the following assuming the fact that, $ob_i, un_j \in R_1$

$$\alpha_{ij} = \frac{d_{ij} \cdot \rho(||ob_i - un^j||)}{\sum_{i \in R_1} d_{ij} \cdot \rho(||ob_i - un^j||)} \quad (15)$$

In the Equation (15) $d_{ij}$ is the degree of seperation established upon the HD and the degree of departure of two PDFs defined on the SRF in the following Equation (16)

$$d_{ij} = \epsilon + \exp-\left(\left[\int_0^1 |f_{ob_i}(y|x_1,x_2) - f_{un^j}(y|x_1,x_2)|^p\right]^{1/p} dy\right) \cdot$$

$$\exp-\left(\sin^{-1}\left[\sqrt{\sin^2\left(\frac{x_{1,ob_i} - x_{1,un^j}}{2}\right) + \cos(x_{1,ob_i}) \cdot \cos(x_{2,un^j}) \cdot \sin^2\left(\frac{x_{2,ob_i} - x_{2,un^j}}{2}\right)}\right]\right) (16)$$

In the Equation (16) $\epsilon$ is included here for the computational adjustment [ (Machuca-Mory, 2013)] along the boundary points of each SR; $\exp-\left(\left[\int_0^1 |f_{ob_i}(y|x_1,x_2) - f_{un^j}(y|x_1,x_2)|^p\right]^{1/p} dy\right)$ specifying modified gaussian distance kernel and here, as a distance we apply the degree of separation between two conditional copula based spatial probability density function (CCSPDF) to capture the probabilistic spatial dissimilarity, and the last part is HD. For $\rho(||ob_i - un^j||)$ in the Equation (15) choice of suitable covariance function is necessary. Therefore, we choose the suitable covariance function among well-defined variogram clouds, for example, Exponential, Gaussian and Spherical, [ (Cressie, 2015)] etc. Applying this concept we adopt the Algorithm (1) to interpolate over the entire spatial surface:

| **Algorithm1: Algorithm of SC Interpolation** |
|---|
| $0 < m \leq 2^k - 1; R_i \cap R_j = \phi$ |
| $S \leftarrow \bigcup_{i=1}^k N_i = \bigcup_{j=1}^m R_j$ |
| $Gen_j = \{(lon_j, lat_j)\}$                                                  # Set of randomly generated points |
| **For** each $j \in Gen_j$ **do** |
| $\vec{v} \leftarrow Presence(Gen_j)$                      #$Presence(.)$ is a binary vector like $f(.)$ in Equation(11) |
|    **IF** $freq(\vec{v}) = 1$    **Then**                                      # $freq(.)$ returns the sum of $\vec{v}$ |
|       $index \leftarrow Index(\vec{v})$                                      #$Index(.)$ returns position of 1 in $\vec{v}$ |
|       $\{ob_1, ob_2, \ldots, ob_r\} \leftarrow$ observed location in $index^{th}$ HSC |
|       $SR(j) \leftarrow$ Choose $p^{th}$ closest SR from $\{ob_1, ob_2, \ldots, ob_r\}$ close to $(lon_j, lat_j)$ |
|       $SC_j \leftarrow \sum_{i \in SR(j)} \alpha_{ij} argmax_y f_{ob_j}(y|lon, lat)$ |
|    **Else** |
|       $\vec{Ind} \leftarrow Index(\vec{v})$ |
|     **For** each $q \in \vec{Ind}$   **do** |
|         $\{ob_1, ob_2, \ldots, ob_r\} \leftarrow$ observed location in $q^{th}$ HSC |
|         $S \leftarrow S.append(\{ob_1, ob_2, \ldots, ob_r\}$       # $X.append(.)$ adds the argument to existing $X$ values |
|     **End For** |
|     $SR(j) \leftarrow Unique(S)$       #$Unique(.)$ removes the duplicate elements from its argument |
|     $SR(j) \leftarrow$ Choose $p^{th}$ closest SR from $\{ob_1, ob_2, \ldots, ob_r\}$ close to $(lon_j, lat_j)$ |
|     $SC_j \leftarrow \sum_{i \in SR(j)} \alpha_{ij} argmax_y f_{ob_j}(y\|lon, lat)$ |
|   **End IF** |
| **End For** |

### 2.3.2 Spatial Bayesian Vine-Copula Estimation

In this section, we introduce spatial vine copula estimation based upon the Bayesian statistical approach (SBVC). Under the square error loss function employing MHA we do the posterior estimate of the parameter in the following way:

$$\pi(\vec{\theta}) = \frac{f_{X_1,X_2|X_3,X_4}(x_1,x_2,x_3,x_4|\vec{\theta}) \cdot p(\vec{\theta})}{\int f_{X_1,X_2|X_3,X_4}(x_1,x_2,x_3,x_4|\vec{\theta}) \cdot p(\vec{\theta}) d\vec{\theta}} \quad (17)$$

In the Equation (17) $f(.|.)$ defines the conditional copula based PDF (CCPDF) derived making use the inherited concept of Figure (1), $p(\vec{\theta})$ denotes the prior PDF of $\vec{\theta}$ and $\pi(\vec{\theta})$ defines the posterior PDF (PPDF) of $\vec{\theta}$. Using the PPDF we'll calculate the posterior estimation of $\vec{\theta}$ under the absolute error loss function. After getting the most updated values $\vec{\theta}$ using MHA we find out the conditional bayesian prediction of two variables in the following:

$$E(\hat{\vec{\theta}}|X_1, X_2, X_3, X_4) = \int \vec{\theta} \cdot \pi(\vec{\theta}) d\vec{\theta}$$
$$E(X_1, X_2|X_3, X_4) = \int \int x_1 \cdot x_2 \cdot f_{X_1,X_2|X_3,X_4}(x_1, x_2, x_3, x_4|\hat{\vec{\theta}}) dx_1 dx_2 \quad (18)$$

Using this Equation (18) we can easily spatially interpolate the target variables on target locations. Here, we consider two SRPs those are, $Y_1(x_1, x_2)$ and $Y_2(x_1, x_2)$ and apply the concept of tail-dependency of a bi-variate copula to measure their hidden reliance. Utilizing VC we find the CCPDF i.e., $F(y_1, y_2|x_1, x_2; \vec{\theta}) = P[Y_1(x_1, x_2) \leq y_1, Y_2(x_1, x_2) \leq y_2|X_1 = x_1, X_2 = x_2; \vec{\theta}]$. Regarding parameter estimation, during fitting MDF, we use UMVUE, EM etc, but to estimate the parameter of the copula family we only consider the posterior estimate. Then using the conditional expectation technique we estimate the $Y_1$ & $Y_2$ values all the randomly generated gridded points and interpolate the values.

### 2.4 Model validation

The accuracy of the models are validated by the following three methods where, $Y(\vec{s}_i)$ is the observed data points and, $\hat{Y}(\vec{s}_i)$ is the predicted value:

1. Mean of Absolute Error (MAE)

$$MAE = \frac{\sum_{i=1}^{n}|Y(\vec{s}_i) - \hat{Y}(\vec{s}_i)|}{n} \quad (19)$$

2. Root of Mean Square Error (RMSE)

$$RMSE = \sqrt{\frac{\sum_{i=1}^{n}(Y(\vec{s}_i) - \hat{Y}(\vec{s}_i))^2}{n}} \quad (20)$$

3. K-fold CV

Then, using Equation (20), (19) we measure the accuracy of the proposed model. K-fold CV is carried out assuming K as 10. The 10-folded CV indicates that the data set is divided into the 10 random sub-sets, and among these data sets, 9 sub-sets are taken as training data set, and the rest 1 is taken as a test data set which is termed as one-leave-one out CV (OLOCV). It helps compare the MAE of proposed and old models

### 3  Study area and Data

To demonstrate the SC, SBVC, and to compare with OK we take Delhi-air pollution as a circumstance study. Delhi, the capital of India is the most polluted due to, rapid urbanization, boosting amounts of traffic, increasing population, and energy consumption at an alarming level. Sometimes, the level of PM $_{2.5}$ concentration in air has reached up to 999 $\mu g/m^3$ [ (Mukherjee, et al., 2018)] and, among all other air pollutants, it affects public health [ (Zheng, Pozzer, Cao, & Lelieveld, 2015)] badly. Boosting levels of automobiles, cars, etc cause higher pollutant concentrations in the air [ (Samal, Gupta, Pathania, Mohan, & Suresh, 2013)]. We look at the air pollution data collected by the monitoring stations, maintained by the Central Pollution Control Board (CPCB), Delhi Pollution Control Committee (DPCC), and the Indian Institute of Tropical Meteorology (IITM). To get the research goal, we collected data on several air pollutants, such as PM $_{2.5}$, PM $_{10}$, NO, NO $_2$, NO $_x$ and wind direction (WD), from the CPCB websites. To map the Spatio-temporal distribution of air quality and deduce the effect of WD on the air pollution in Delhi, these data play an important role. The data were collected over 24 hours, and the period was taken from $1^{st}$ February 2017 to $31^{st}$ December 2021. The Figure (4) depicts the temporal variability of daily PM $_{2.5}$ emission which is cyclic in nature after a fixed time stand. There is always a higher concentration witnessed from almost the end of November to the end of December (Figure (4)) around $400\mu g/m^3$ and sometimes it grows up to $800\mu g/m^3$ which is very much alarming for the human life, primarily during winter due to burning of firecrackers, agricultural crop burning, etc.

**PM $_{2.5}$ concentration over the whole spatial surface**

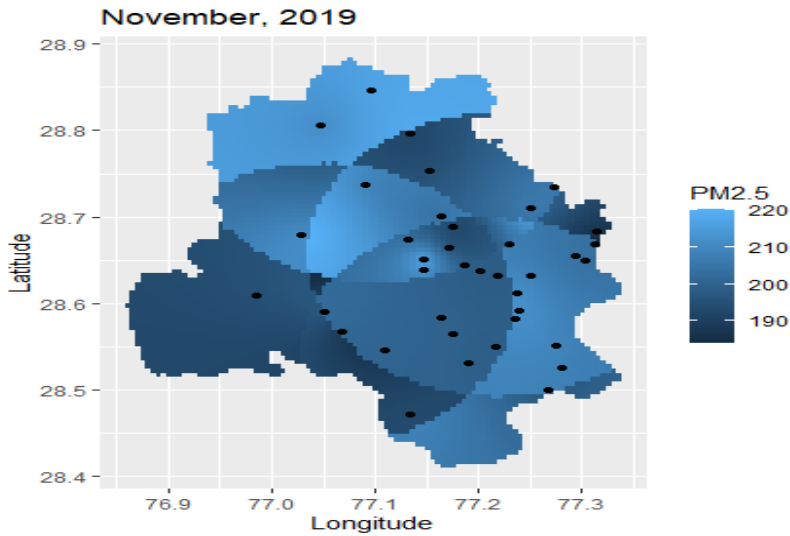

**Figure  3:** The interpolated PM $_{2.5}$ values of November in the year, 2019

There are 38 monitoring stations in this data set, as shown in Figure (2 B). According to Figure (3), we detect that the Northern part of Delhi is very much sensitive to pollution whereas the Central, East, and West parts of Delhi are less sensitive regarding the pollutant concentrations in the air. According to Figure (3), the PM $_{2.5}$ concentration in the Northern part of Delhi can reach up to $220\mu g/m^3$ whereas in the Central, East, and West part of Delhi that is limited into 190 to 200 $\mu g/m^3$. As a result, the Spatio-temporal variability in air pollutant concentrations is visible.

However, there are two shortcomings to applying spatial interpolation techniques to interpolate (i) the Delhi NCR region is far away from other monitoring stations in Delhi, (ii) The missing Data.

**Temporal Variability of $PM_{2.5}$**

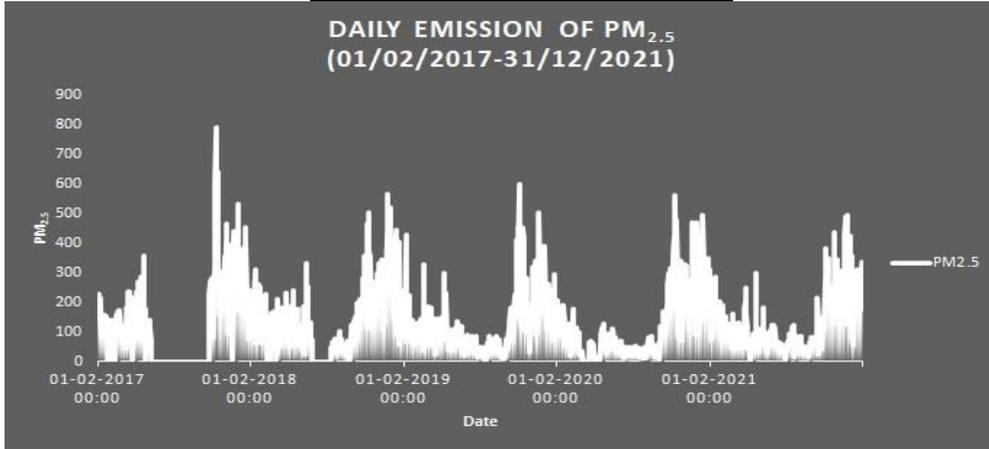

**Figure 4:** The time series plot of daily $PM_{2.5}$ emission from $1^{st}$ Feburary, 2017 to $31^{st}$ December, 2021 in the study area, Delhi.

**Histogram of daily $PM_{2.5}$ emission**

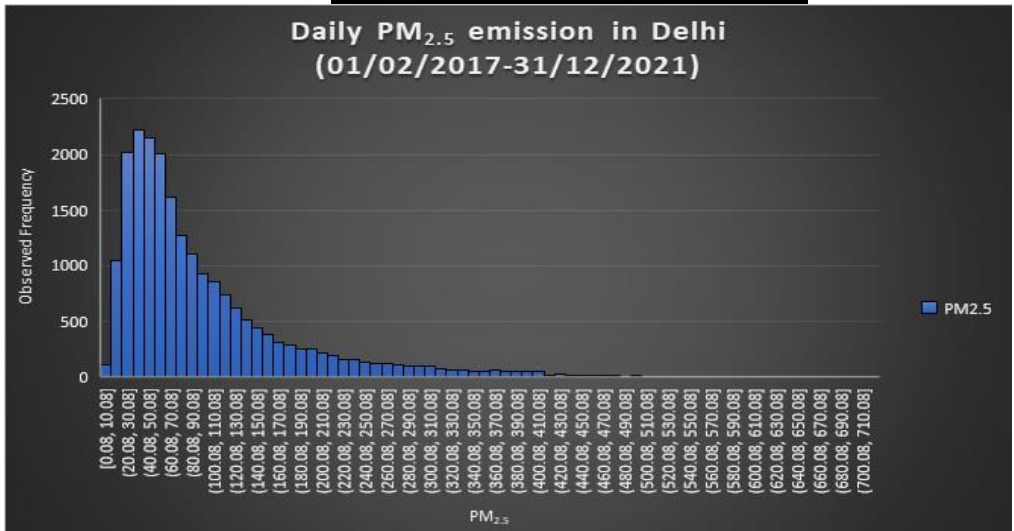

**Figure 5:** The Histogram of daily $PM_{2.5}$ emission from $1^{st}$ Feburary, 2017 to $31^{st}$ December, 2021 in the study area, Delhi.

We can easily conclude from the Figure (5) that the observed frequency distribution of daily $PM_{2.5}$ emission during this period is positively skewed which gives an idea of how to fit the positively skewed distribution such as log-Normal, Gamma, Exponential, Weibull, etc depending upon the tail distribution. Precisely there is a higher concentration in the interval from $30 - 40 \mu g/m^3$. Figure (6) provides a brief overview of the variability and a five-point summary of the pollutants and WD which establishes the fact that $PM_{10}$ and $PM_{2.5}$ have higher variability compared to other pollutants and the variance of WD also sensitive.

**Box-plot of daily emission of pollutans and WD**

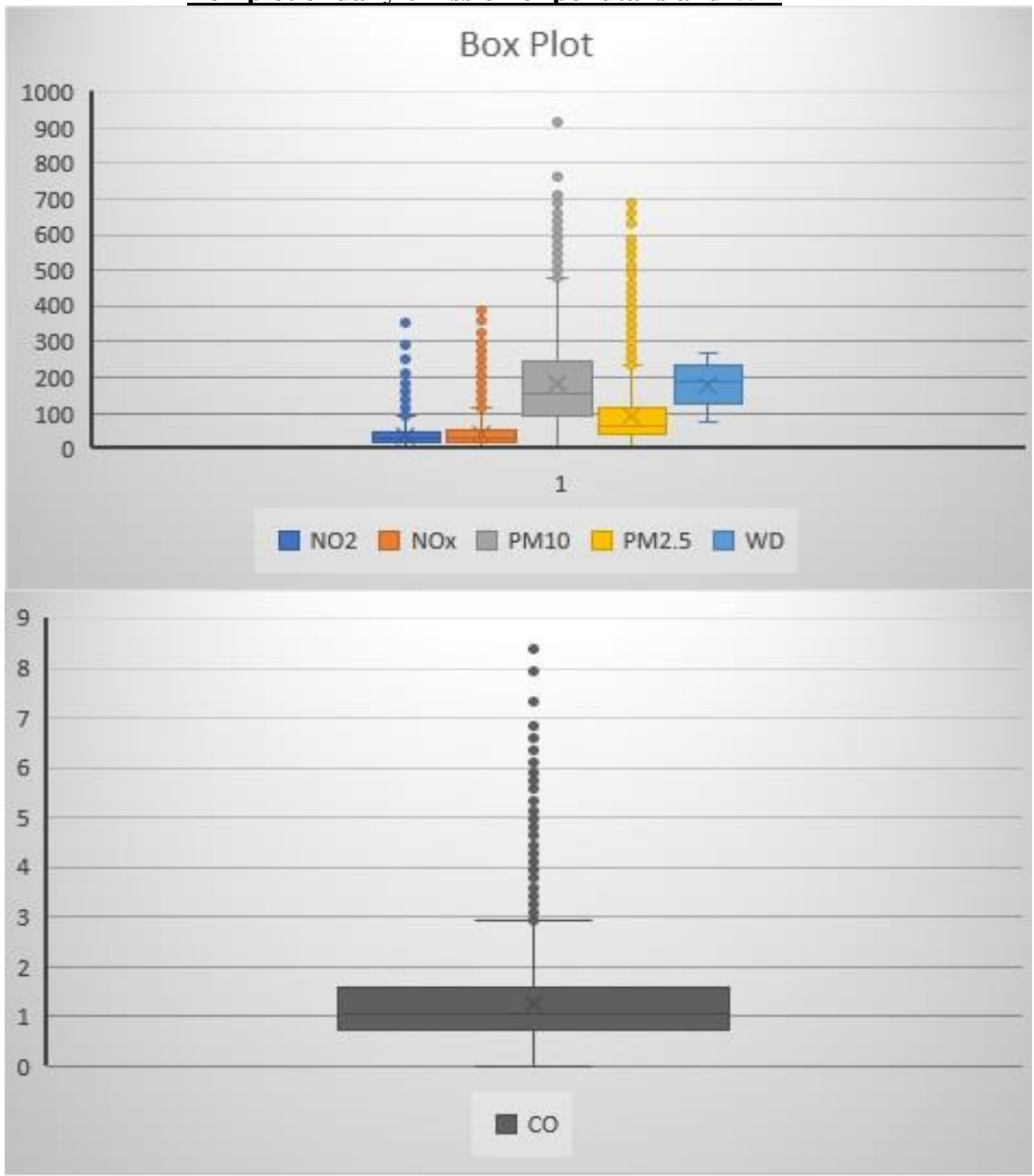

**Figure 6:** The Box-Plot of daily PM $_{2.5}$, PM $_{10}$, NO $_2$, NO $_x$ emission and WD from $1^{st}$ Feburary, 2017 to $31^{st}$ December, 2021 in the study area, Delhi.

## 4 Results and Discussion

This section goes over how to compare two new models, SC and SBVC to other well-known spatial models step by step. Following that, we will attempt to provide a brief overview of pollutant concentrations in the future, as well as discuss how an important meteorological parameter can affect pollution concentrations mathematically.

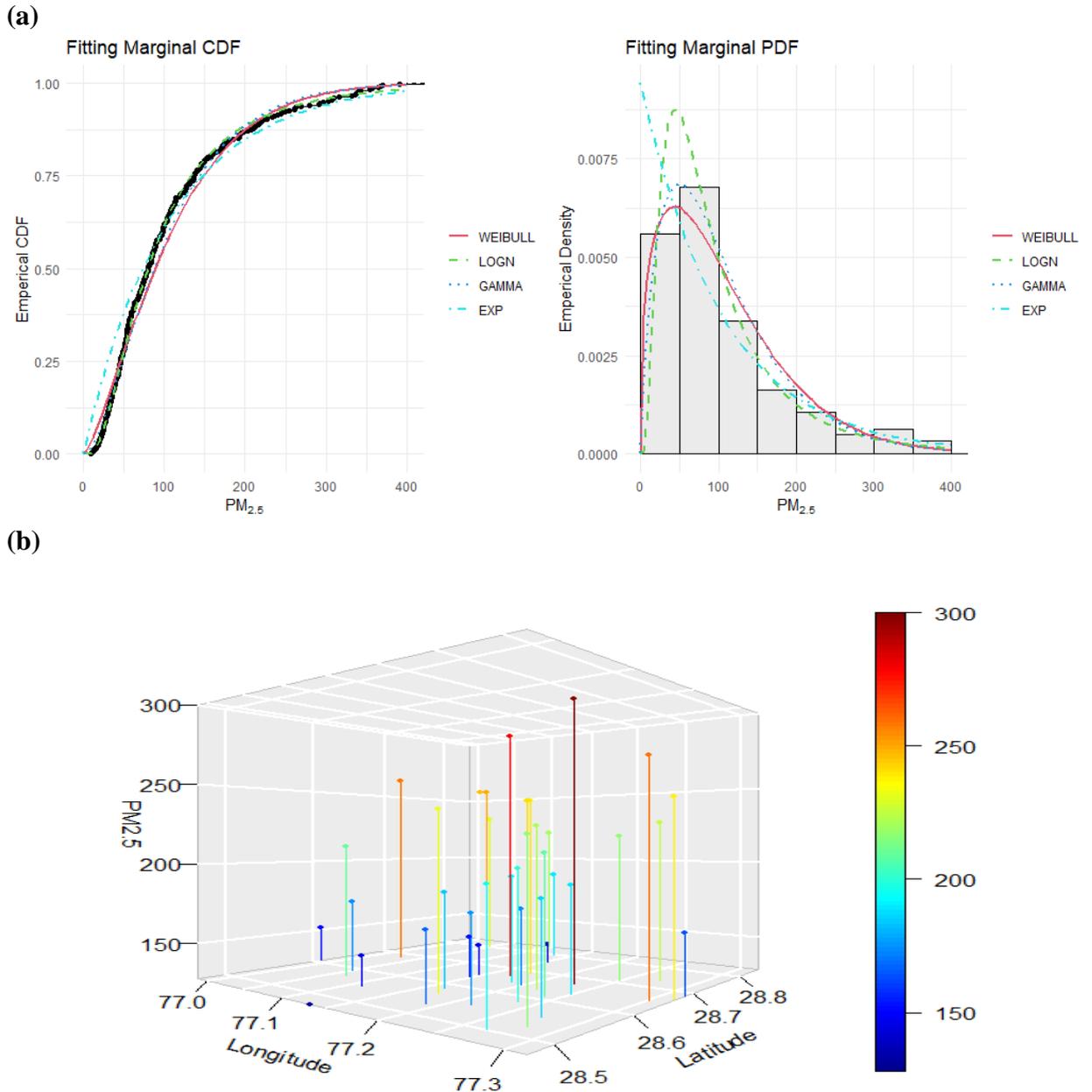

**Figure 7: (a)** The emperical marginal CDF is fitted by the marginal positively skewed parametric CDF and the fitting of marginal PDF. **(b)** The nature of spatial variation of a RV. We fit the parametric marginal CDF and PDF on the empirical CDF and PDF of an RV

based on the AIC, BIC value, and KS test statistic value in Figure (7 a). Because the empirical PDF is positively skewed, we only consider well-known positively skewed distributions such as Weibull, Log-normal (LN), Gamma, and Exponential distributions, and Table (1) shows that the LN distribution is suitable to fit based on the lowest AIC, BIC, and KS test statistic value. Similarly, we fit the circular family distributions on WD and discover that the VM distribution is the best PDF to fit.

**Table 1: The value of KS statistic, AIC and BIC to determine the feasible marginal parametric PDF**

| Test Criteria | Weibull | Log-normal | Gamma | Exponential |
|---|---|---|---|---|
| KS | 0.06884229 | 0.02849193 | 0.06276518 | 0.1478233 |
| AIC | 365.360 | 322.296 | 346.187 | 413.547 |
| BIC | 373.098 | 330.035 | 353.926 | 417.416 |

The next step is to estimate the parameter of the MDF. We already discussed the disadvantages of using MLE to estimate the parameter in Section (2.1). As a result, we can use the EM algorithm to obtain the updated shape and scale parameters of the LN distribution. We discuss how the LogLik value converges to a fixed value after a certain number of iterations in Figure (8). The required number of iterations for the EM algorithm in this case study is 223, after which the difference between the two LogLik values is negligible. Figure (7 b) depicts how the PM $_{2.5}$ value varies with respect to latitude and longitude. While the Longitude (Lon) ranges from 77.0 to 77.1 and the Latitude (Lat) varies, from 28.5 to 28.6, the PM $_{2.5}$ concentration is generally within $100 - 150 \mu g/m^3$ but if Lon varies from 77.15 to 77.3, the PM $_{2.5}$ concentration becomes high and it ranges from $150 - 200 \mu g/m^3$. Similarly, while Lat is varying from $28.6 - 28.7$ then the most spatial variability of PM $_{2.5}$ is detected in every interval of Lon and sometimes reaches up to $300 \mu g/m^3$ while the Lat varies from $28.7 - 28.8$ the spatial variation is identified and the variation of PM $_{2.5}$ is almost lying between $200 - 250 \mu g/m^3$.

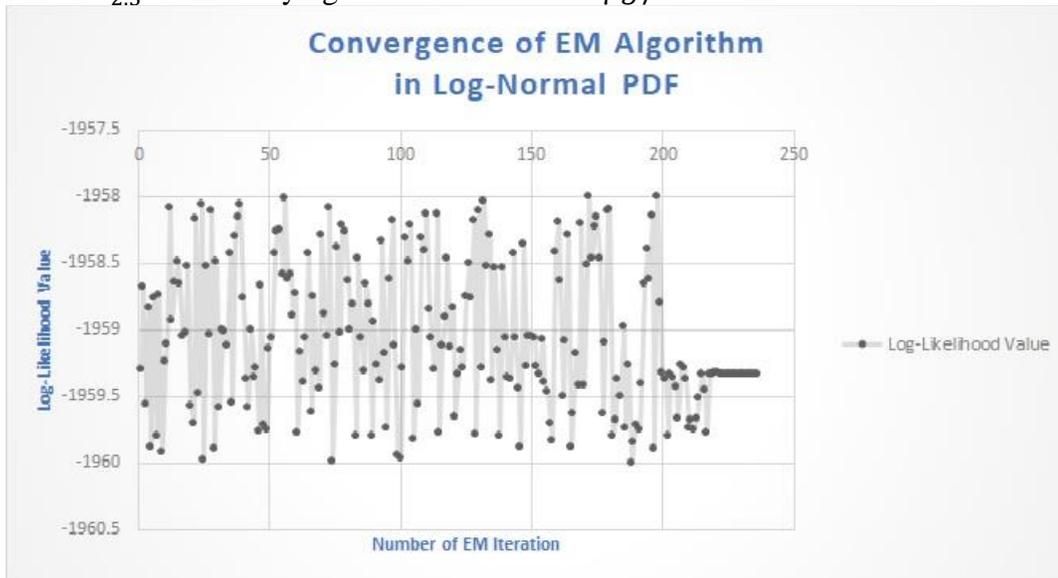

**Figure 8:** The convergence of the Log-Likelihood value after updating the value of the parameters in each iteration using EM algorithm.

However, during fitting VM distribution we use the concept of UMVUE which is mentioned in the Theorem (2.1) in the Section (2.1) to get the shape and scale parameter of the VM distribution with better accuracy. In the following Table (2) we discuss the shape and scale parameters of LN and VM PDF and corresponding the last updated LogLik values.

**Table 2: Details and updated values of shape and scale parameter and the corresponding Log-likelihood values**

| PDF | Shape | Scale | LogLik |
|---|---|---|---|
| LN | 4.3764856 | 0.7701984 | -1959.331124 |
| VM | 3.583 | 1.908 | -32.41559 |

Our goal now is to run the two novel spatial interpolation algorithms mentioned in Sections (2.3.1) and Section (2.3.2) and compare them to other spatial interpolation approaches. Using the threshold criteria mentioned in Section (2.3.1) we divide the entire spatial domain into 4 HSC and consider the cutoff radius is 18026m as shown in Figure (9 a).

**Optimal number of HSC and optimal HSC size**

**(a)**

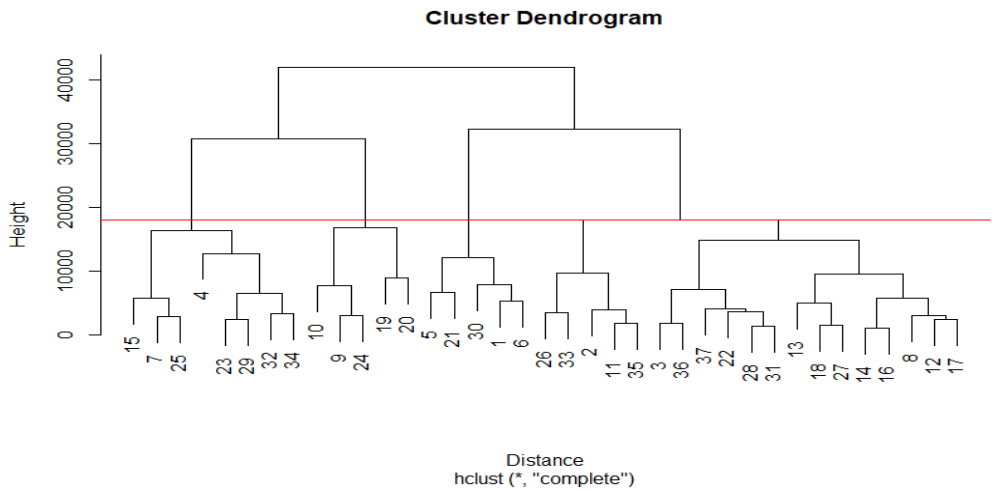

**(b)**

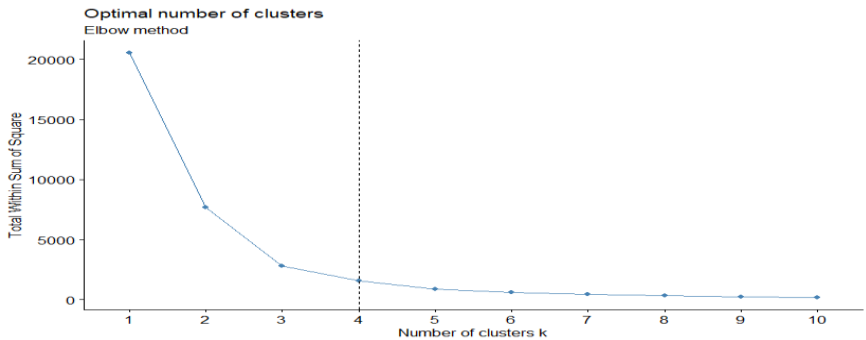

**Figure 9:** Optimal number of spatial HSC showed in **(a),** and in **(b)** we present its optimum HD.

In Figure (9 (A)) of the cluster dendrogram, the height represents the HD, and in Figure (9 (B)) we plot SSW along the Y-axis and the optimal number of HSC along the X-axis. The following discussion focuses on the tail dependence of two RVs, as shown in Figure (10). These two RVs in this case study are PM $_{2.5}$ and WD.

### Tail-dependency and Joint-CDF

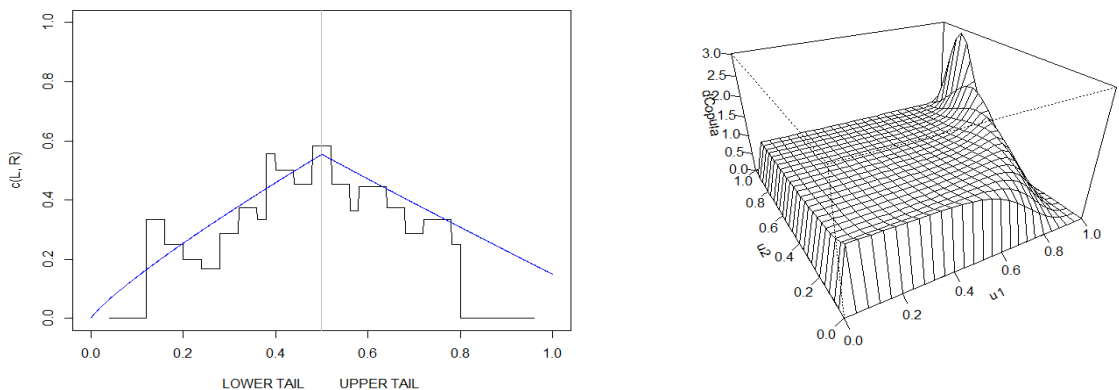

**Figure 10:** In the (*left*) part the discussion regarding the lower tail and upper tail dependency of two RV and in the (*right*) part the joint CDF of two RV.

The upper tail dependence and lower tail dependence between two RVs, X and Y is defined in the following:

$$\begin{aligned} d_u &= \lim_{u \to 1} P\big[Y \geq F_y^{-1}(u) | X \geq F_x^{-1}(u)\big] = \lim_{u \to 1} \frac{C(1-u, 1-u)}{(1-u)} \\ d_l &= \lim_{u \to 0} P\big[Y \leq F_y^{-1}(u) | X \leq F_x^{-1}(u)\big] = \lim_{u \to 0} \frac{C(u,u)}{u} \end{aligned} \quad (21)$$

The Equation (21) shows the behaviour of one RV while othe one converging to the extreme values. We can conclude from this Figure (10) *left* that after 0.8 the upper tail of their distributions is independent and lesser than 0.1, the lower tail of their distributions is independent. As a result, we can say that higher values of PM $_{2.5}$ concentration are unaffected by WD because there is a very low concentration at that point but where the marginal PDF of PM $_{2.5}$ is moderate, there is a significant tail dependence on WD. The joint CDF of PM $_{2.5}$ and WD is plotted in the Figure (10) on *right* applying $BiCopSelect()$ function in R, where the fitted copula is Rotated Twan Type-2 Copula with estimated Kendall's $\tau = 0.1341$ and the LogLik value is $-1.204$ which is the highest of any copula family, including GC, t-Copula, Frank, Clayton, Joe, and so on. We want to use our novel copula-based spatial interpolation algorithm (SC) in a Bayesian framework after fitting the copula using $CDVineCondFit()$ function in R. The posterior distribution and posterior estimate of the parameter are critical in this context. MHA is used in this context to obtain the posterior estimate of the parameters.

### Rate of convergence of MHA Algorithm

**(a)**

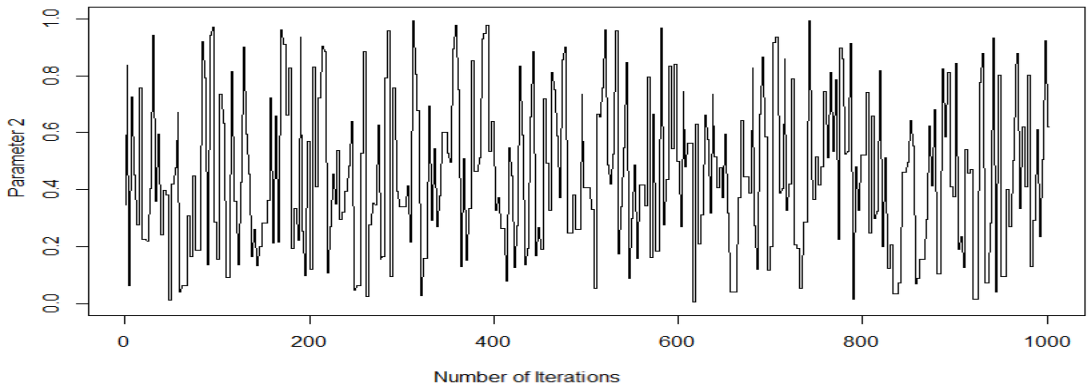

**(b)**

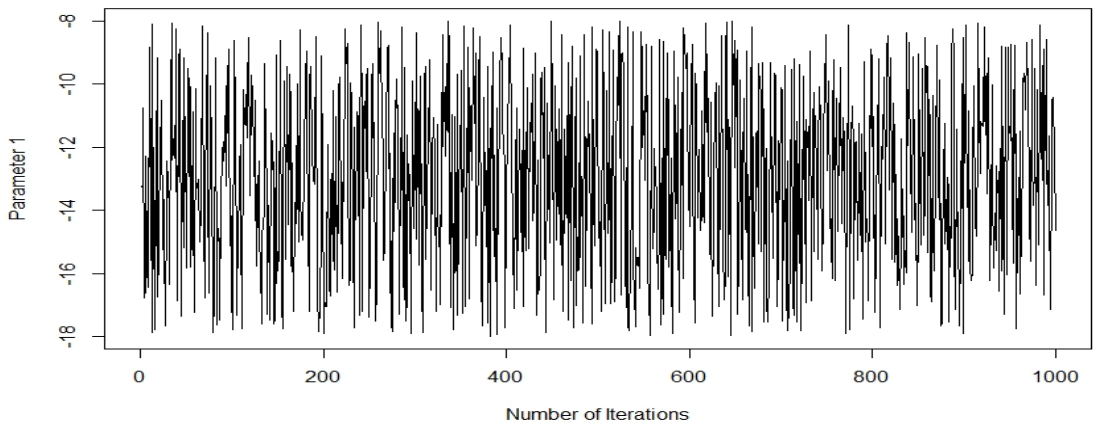

**Figure 11:** The rate of convergence of the two parameters of Rotated Tawn Type-2 copula family. In (**b**) the first parameter of the copula family and in (**a**) the second parameter of the copula family are estimated.

According to this Figure (11) we use the concept of Bayesian Inference to give the posterior estimate, assuming that the parameter prior distributions are uniform and truncated normal distributions. Following that, we use MHA to obtain the posterior estimate under the MSE loss function, which is 0.04898261 (from Figure (11 a)) and −13.61893 (from Figure (11 b)), respectively. The rate of convergence of two parameters is plotted in the Figure (11), depicting that the rate of convergence of parameter 1 is faster than that of Parameter 2. Now we will look at how WD and spatial clustering affect the variance of PM $_{2.5}$ in Table (3) and the Figure (15).

**Table 3:** Two-way ANOVA to explain the dependence of PM $_{2.5}$ emission on WD and SC

| Treatment | Df | SS | MS | F Ratio | P-Value |
|---|---|---|---|---|---|
| WD | 21 | 20599.2864 | 980.9184 | 3.696 | 0.02404* |
| Cluster | 3 | 3623 | 1207.8 | 4.550 | 0.0334* |
| Cluster· WD | 3 | 1283 | 427.7 | 1.611 | 0.2543 |
| Residuals | 9 | 2389 | 265.4 | | |

Signif. codes: 0 '***' 0.001 '**' 0.01 '*' 0.05 '.' 0.1 ' ' 1

As a result, in the two-way analysis of the variance model (Two way ANOVA), we consider PM $_{2.5}$ as a dependent variable and WD and clusters as independent variables. Along with the columns of Table (3), we represent Treatments, Degrees of freedom (Df), Sum of square (SS), Mean Square (MS), F-Ratio, and P-value, and along the rows, we represent WD, Cluster, their interaction effect, and residuals. We can see from Table (3) that there is a significant impact of WD and clustering on PM $_{2.5}$ emission at the 0.05 level of significance. However, the interaction effect of WD and Clusters has no significant impact on PM $_{2.5}$ emission. To aid comprehension, we present a graphical representation of these ANOVA tables in Figure (15) in Section (6.3) where WD is represented along the X-axis, PM $_{2.5}$ is represented along the Y-axis, and each spatial cluster is used as a panel. In the SC interpolation method, we investigate another factor, spatial ACF, which is employed as an important weight to counteract spatial variability across all lags.

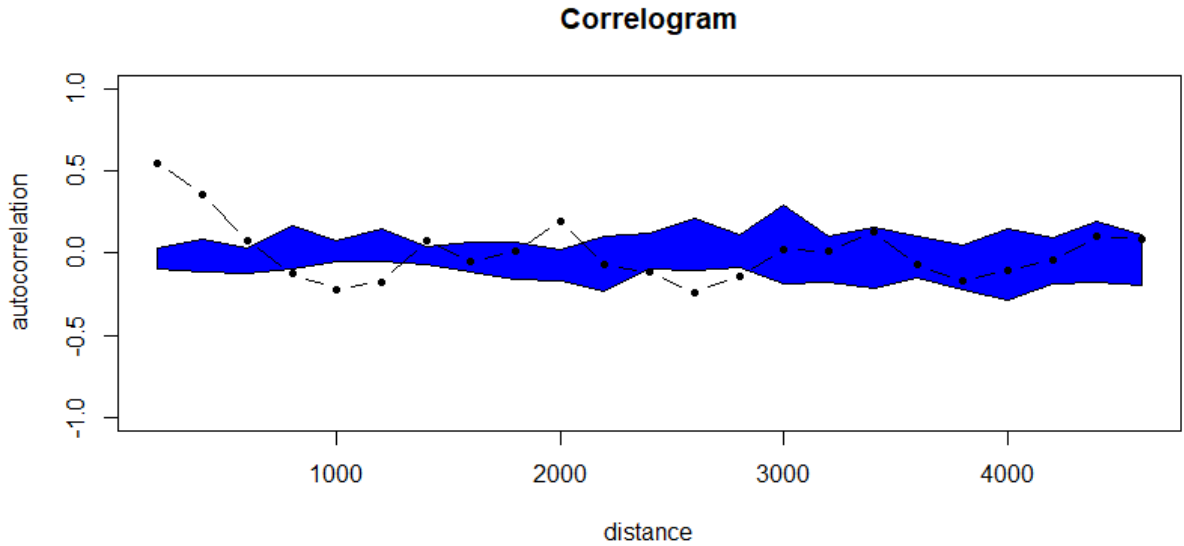

**Figure 12:** Spatial ACF corresponding to every spatial lag, we plot the lag distance along the X-axis and the ACF along the Y-axis.

As a result, in the Figure (12), we investigate how the ACF varies with respect to the spatial lag. In this Figure (12), we notice that the value of ACF is comparatively higher for nearby stations than for stations further away, and we use the blue shaded region to give a brief idea of the interval of variation of ACF values. In this case study the fitted variogram model is Matern variogram model with nugget: 0; sill: 617; range: 0.02 and kappa: 0.09. Utilizing this value and the other distance weights in the Equation (14) we calculate CCPDF in every unobserved location. In this case study, we assume that $\epsilon$ is 0.4224 and $p = 2$ in the Equation (14).

The entire framework is now ready to execute the new spatial copula interpolation (SC) described in Section (2.3.1) and Bayesian Spatial-Vine Copula (SBVC) described in Section (2.3.2). As a result, we create an SRF within each HSC and focus on the spatial region between them. For SC, we assume that Lat and Lon have a bi-variate uniform distribution, and PM $_{2.5}$ has a LN distribution, and the suitable copula is Clayton Copula among other copula familes like, Gauusian, t-copula, archimedian-copulas, based on AIC, BIC, and LogLik values, to find their joint CDF using $mvdc()$ function in R, with a parameter of 0.01697 and a dimension of 3. Then, using the Equation (22), we obtain the required CCDF.

$$F(y_1|x_1,x_2)$$

$$= \frac{k_1 \cdot (x_1 x_2)^{-k_2} \cdot \int_{-\infty}^{y_1} \left(1 + (\frac{1}{x_1^\theta} + \frac{1}{x_2^\theta} - 1)\right) \cdot (F_{Y_1}(t))^{-k_3} \cdot (F_{Y_1}(t))^{k_3 - k_2} \cdot f_{Y_1}(t) dt}{c(x_1, x_2)} \quad (22)$$

Using the Equation (14), (15), and (16) we get the CCPDF of each unobserved location. Then using the Algorithm (1) we get the interpolated values.

**Spatial Interpolation**

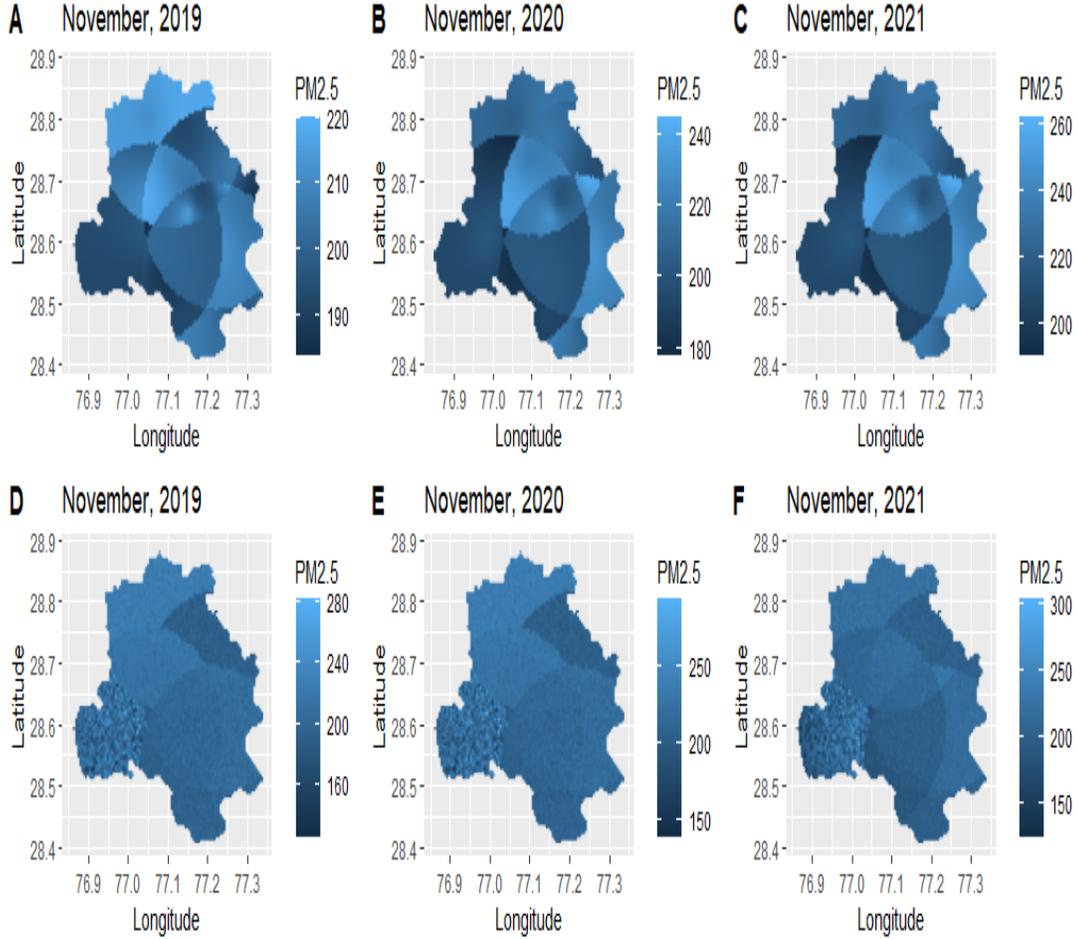

**Figure 13:** Spatial Interpolation of PM $_{2.5}$ during the month of November, in 2019, 2020 and 2021. In *Top* the spatial interpolation technique SC is used and in *below* the SBVC algorithm is used as a spatial interpolation algorithm. Along X-axis we plot Longitude, along Y-axis we plot latitude and along the whole surface we plot the PM $_{2.5}$.

In the Figure (13) we plot the monthly PM $_{2.5}$ emission during the month of November for the three years, 2019, 2020, and 2021. According to this Figure (13) we detect that using SC in November, 2019, the PM $_{2.5}$ emission varies from $180 - 220 \mu g/m^3$ whereas using SBVC ranges from $120 - 280 \mu g/m^3$ (Figure (13)), similar patterns are observed in the year 2020 and 2021 as well. If we carefully examine Figure (13) carefully we can look that the variation of SBVC is greater than that of SC and the Northern and South-East part of Delhi is highly sensitive to pollution. We notice that in the western part of Delhi the SBVC is too effective to show the strong spatial trend, as a result the PM $_{2.5}$ emission is random in nature (Figure (13)). We investigate the relationship between the observed and predicted values of three methods: SC, SBVC, and OK in the Figure (14 (*Top*)), and we discover that there is a strong relationship between the observed and predicted values in SC, followed by SBVC, and finally OK. As a result, we can make the critical observation that the power of explainable variation in SC is greater than SBVC, which is better than OK. It is clear that, the measurement error, MAE and RMSE of SC is lesser than SBVC, and lastly OK as demonstrated by Figure (14 (*Bottom*)).

Although the SC method outperforms the other two, there are some areas where improvements are possible, such as: (i) When clustering, we assume that the rate of inclusion of geo-spatial points in a cluster is constant, but this can vary in practice. (ii) we ignore the effect of extreme values during interpolation; (iii) we use the concept of degree of departure of characteristics between observed and unobserved points and spatial continuity concurrently, but it sometimes contradicts the concept of spatial continuity; and (iv) we do not pay enough attention to the data set's temporal stationarity; (v) to estimate the parameter of copula, applying EM algorithm. SBVC accepts the same drawbacks, but one additional drawback is that when determining spatial trend, but it is sometimes ineffective to explore.

**Quality of interpolation of SC, SBVC, and OK**

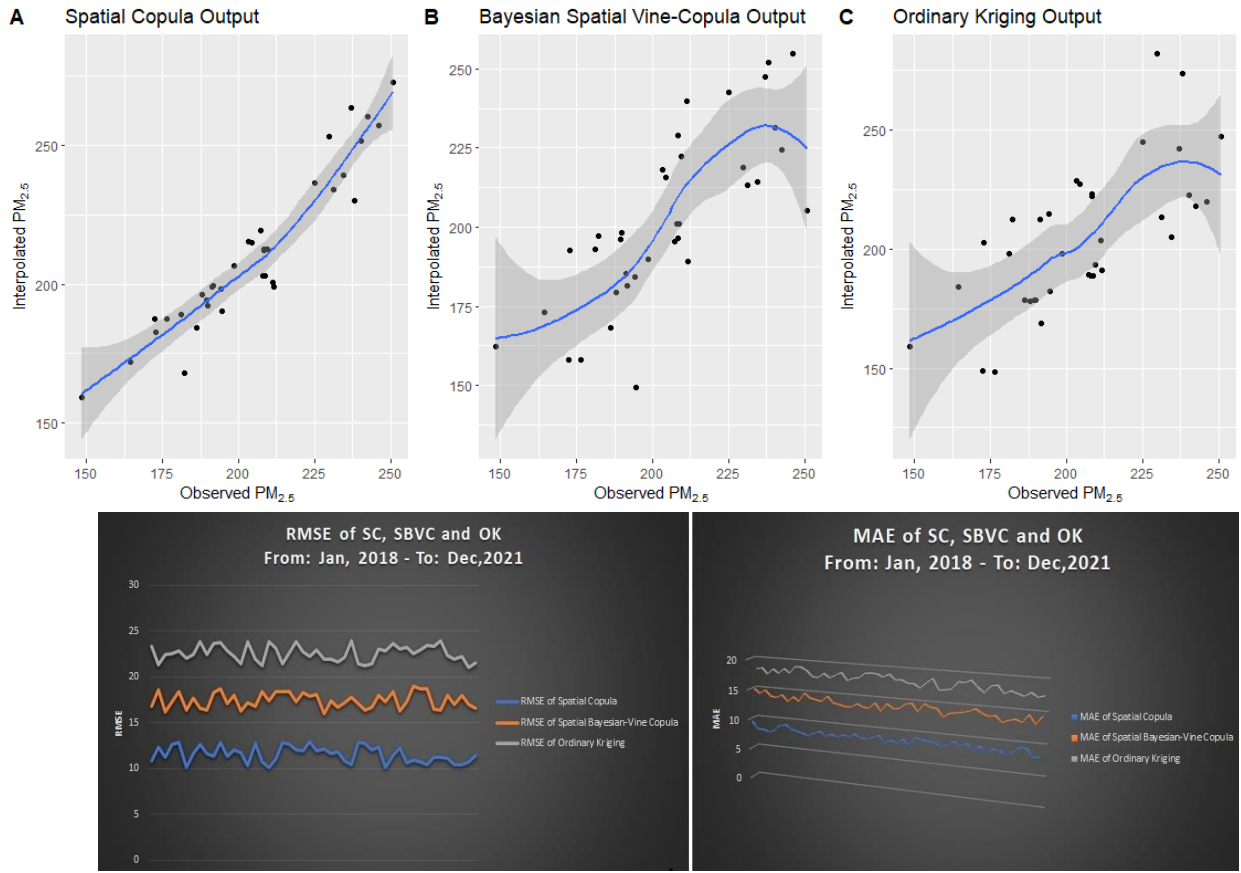

**Figure 14:** (*Top*)Relationship between the observed and predicted values of three methods: SC, SBVC, and OK. (*Bottom*) Comparison of the performance of three methods: SC, SBVC, and OK. Along Y-axis we plot the RMSE and MAE of the four years from January, 2018 to December, 2021.

## 5 Conclusion

The introduced SC and SBVC are extensions of the previous spatial copula-based models that address issues such as bin selection, the use of MLE to estimate the parameter in missing data sets, and so on. When compared to other geo-statistical models, the entire SC and SBVC are very effective and provide nearly accurate results (from Figure (14 (*Bottom*))). This SC model produces better results for specially skewed spatial random fields and provides a mathematical argument for selecting important covariates. This study provides an idea of alternative distance weights and distance functions that are very effective in capturing spatial variation. A temporal extension of this algorithm is possible, which motivates further research. This model is explained in this study using a real-world data set of PM $_{2.5}$ concentrations in the air, but this algorithm can be used in other scenarios such as mining, temperature modeling, meteorological modeling, and so on. This algorithm may be more advantageous than other spatial estimation models because it makes no assumptions about gaussian distribution, stationarity, dynamic behavior, or skewed data sets.

## 6 Appendix

### 6.1 Proof of Theorem (2.1)

*Proof.* If $X \sim VM(k, \mu)$ then we know the corresponding charachteristic function of $X$ is $\phi_n(x) = E[e^{inx}]$

$$E[e^{inx}] = \int_0^{2\pi} \frac{e^{inx} \cdot e^{k \cdot \cos(x-\mu)}}{2\pi I_0(k)} dx \qquad (23)$$
$$= \frac{I_{|n|}(k) \cdot e^{in\mu}}{I_0(k)}$$

In the Equation (23) the term $I_n(k) = \frac{\int_0^\pi e^{k \cdot \cos(x)} \cos(nx) dx}{\pi}$. In Equation (23) putting $n = 1$ we get,

$$E(e^{ix}) = \frac{I_1(k) \cdot e^{i\mu}}{I_0(k)} \qquad (24)$$

and putting $n = -1$ we get,

$$E(e^{-ix}) = \frac{I_1(k) \cdot e^{-i\mu}}{I_0(k)} \qquad (25)$$

Adding and subtracting Equation (24) and Equation (25) we get

$$E\left(\frac{I_0(k) \cdot \cos(x)}{I_1(k)}\right) = \cos\mu$$
$$E\left(\frac{I_0(k) \cdot \sin(x)}{I_1(k)}\right) = \sin\mu \qquad (26)$$

Therefore, from the Equation (26) the statistic $T_1(x) = \frac{I_0(k) \cdot \cos(x)}{I_1(k)}$ and $T_2(x) = \frac{I_0(k) \cdot \sin(x)}{I_1(k)}$ are the unbiased estimators of $\cos\mu$ and $\sin\mu$ respectively.

Here, the PDF of $X$ is denoted as $f(x)$ and the parameter space is defined as $\Phi$ and support is defined as $\mathcal{X}$.

$$f(x) = \frac{e^{k \cdot \cos(x-\mu)}}{2\pi \cdot I_0(k)}$$
$$= exp[k \cdot \cos(x - \mu) - \log(2\pi I_0(k))] \qquad (27)$$
$$= exp[k \cdot \cos(x)\cos(\mu) + k \cdot \sin(x)\sin(\mu) - \log(2\pi) - \log(I_0(k))]$$

From the Equation (27) we write the likelihood function as product of two terms moreover this VM distribution satisfying the following properties:

1. $\mathcal{X}$ is $[0, 2\pi]$ therefore, it is independent upon the parameter.
2. $\Phi = \{(\mu, k): \mu \in \mathcal{R}; k > 0\}$ which indicating it is an open interval.
3. Here $\{1, cos(x), sin(x)\}$ and $\{1, cos(\mu), sin(\mu)\}$ are Linearly Independent (LIN).

Therefore, we tell that the PDF is belonging Two-PEF. Therefore, $cos(x)$ and $sin(x)$ are complete and sufficient statistic of $cos(\mu)$ and $sin(\mu)$. Using Lehmann-Scheffe Theorem they are the UMVUE. Moreover, using Equation (23) replacing $n = 2$ we get

$$E[cos2x] = \frac{I_2(k)}{I_0(k)}$$
$$\Rightarrow E(cos^2(x)) = \frac{1}{2} + \frac{I_2(k) \cdot cos(2\mu)}{2I_0(k)} \quad (28)$$
$$\Rightarrow E(sin^2(x)) = \frac{1}{2} - \frac{I_2(k) \cdot sin(2\mu)}{2I_0(k)}$$

Using Equation (28) we get

$$var(cos(x)) = \frac{1}{2} + \frac{I_2(k) \cdot cos(2\mu)}{2I_0(k)} - \left(\frac{I_1(k) \cdot cos(\mu)}{I_0(k)}\right)^2 \quad (29)$$
$$var(sin(x)) = \frac{1}{2} - \frac{I_2(k) \cdot sin(2\mu)}{2I_0(k)} - \left(\frac{I_1(k) \cdot sin(\mu)}{I_0(k)}\right)^2$$

### 6.2 EM algorithm estimation of parameters of VM distribution

Likewise LN distribution $\mathcal{Q}\left((\mu, k)|(\mu^{(m)}, k^{(m)})\right)$ is the updated CElikC at $m^{th}$ iteration is:

$$\mathcal{Q}\left((\mu, k)|(\mu^{(m)}, k^{(m)})\right) = E^c_{(\mu^{(m)}, k^{(m)})} \begin{bmatrix} \sum_{i=1}^{n_2} kcos(w_i - \mu) - n_2 \cdot log(2\pi I_0(k)) \\ \sum_{i=1}^{n_1} kcos(w_i - \mu) - n_1 \cdot log(2\pi I_0(k)) \end{bmatrix} \quad (30)$$

Using the Equation (30) we complete M-step and find the most updated values.

### 6.3 Two-way ANOVA

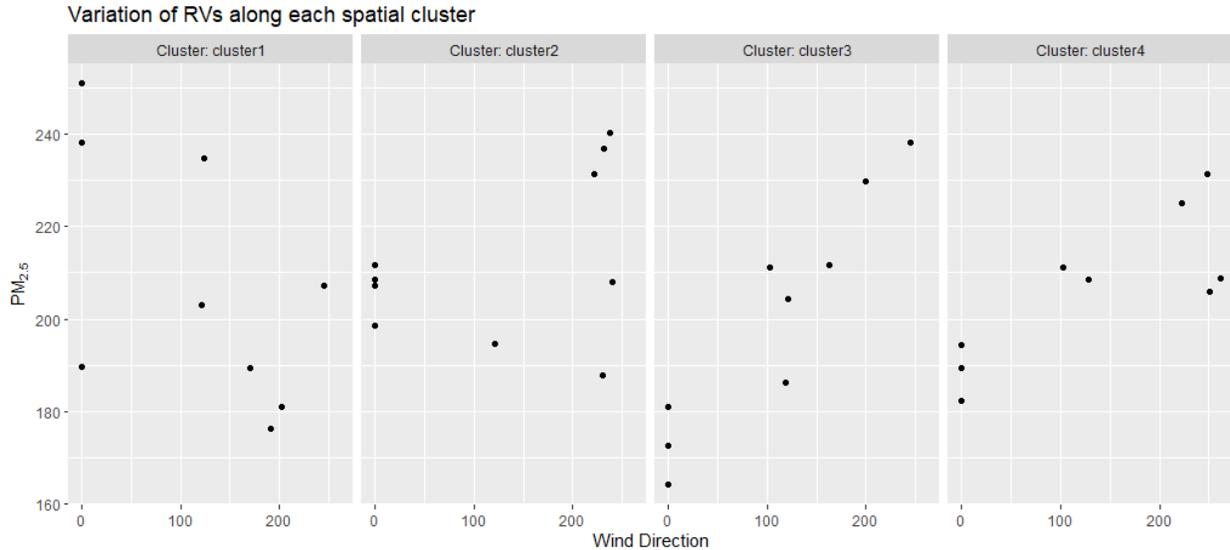

**Figure 15:** How WD and Spatial cluster make an impact on PM $_{2.5}$ in this case study.

## 7 Data Availability

The data will be available upon request to the corresponding author. Otherwise it can be possible to download the data directly can be downloaded from this link "https://app.cpcbccr.com/ccr/#/caaqm-dashboard-all/caaqm-landing/data" .

## 8 Acknowledgments

**Funding:** No funding for this paper.
**Author Contributions:** The three authors have made equal contribution to this paper.
**Competing interests:** Authors state no conflict of interest.